\documentclass{bioinfo2}
\copyrightyear{2011}
\pubyear{2011}

\usepackage{natbib}
\DeclareMathOperator*{\argmax}{argmax}

\begin{document}
\firstpage{1}

\title[Inference using sequencing data]{A statistical framework for SNP calling, mutation discovery,
		association mapping and population genetical parameter estimation from sequencing data}

\author[Li]{Heng Li}

\address{Broad Institute, 7 Cambridge Center, Cambridge, MA 02142, USA}

\maketitle

\begin{abstract}
\section{Motivation:}
Most existing methods for DNA sequence analysis rely on accurate sequences or
genotypes.  However, in applications of the next-generation sequencing (NGS),
accurate genotypes may not be easily obtained (e.g. multi-sample
low-coverage sequencing or somatic mutation discovery).
These applications press for the development of new methods for analyzing
sequence data with uncertainty.
\section{Results:}
We present a statistical framework for calling SNPs, discovering somatic
mutations, inferring population genetical parameters, and performing
association tests directly based on sequencing data without explicit genotyping
or linkage-based imputation. On real data, we demonstrate that our method
achieves comparable accuracy to alternative methods for estimating site allele
count, for inferring allele frequency spectrum and for association mapping. We
also highlight the necessity of using symmetric data sets for finding somatic
mutations and confirm that for discovering rare events, mismapping is
frequently the leading source of errors.
\section{Availability:} http://samtools.sourceforge.net
\section{Contact:} hengli@broadinstitute.org
\end{abstract}

\section{INTRODUCTION}

The 1000 Genomes Project~\citep{1000-Genomes-Project-Consortium:2010qc} sets an
excellent example on how to design a sequencing project to get the maximum
output pertinent to human populations. An important lesson from this project is to
sequence many human samples at relatively low coverage instead of a few samples
at high coverage. We adopt this strategy because with higher coverage, we will mostly
reconfirm information from other reads, but with more samples, we will be able to
reduce the sampling fluctuations, gain power on variants present in multiple
samples and get access to many more rare variants. On the other hand,
sequencing errors counteract the power in variant calling, which necessitates a
minimum coverage.  The optimal balancing point is broadly regarded to be in the
2--6 fold range per sample~\citep{Le:2010uq,Li:2011fk}, depending on the
sequencing error rate, level of linkage disequilibrium (LD) and the purpose of the
project.

A major concern with this design is that at 2--6 fold coverage per sample,
non-reference alleles may not always be covered by sequence reads, especially
at heterozygous loci. Calling variants from each individual and then combining
the calls usually yield poor results. The preferred strategy is to enhance the
power of variant discovery by jointly considering all
samples~\citep{Le:2010uq,Li:2011fk,Depristo:2011vn,Nielsen:2011fk}. This
strategy largely solves the variant discovery problem, but acquiring accurate
genotypes for each individual remains unsolved. Without accurate genotypes,
most of previous methods (e.g. testing Hardy-Weinberg equilibrium (HWE)
and association mapping) would not work.

To reuse the rich methods developed for genotyping data, the 1000 Genomes
Project proposes to impute genotypes utilizing LD across
loci~\citep{Li:2009gb,Browning:2009jl,Howie:2009mb,Li:2010ky}.  Suppose at a
site $A$, one sample has a low coverage. If some samples at $A$ have high
coverage and there exists a site $B$ that is linked with $A$ and has sufficient
sequence support, we can transfer information across sites and between
individuals, and thus make a reliable inference for the low-coverage sample at
$A$. The overall genotype accuracy can be greatly improved.

However, imputation is not without potential concerns. Firstly, imputation
cannot be used to infer the regional allele frequency spectrum (AFS) because
imputation as of now can only be applied to candidate variant sites, while we
need to consider non-variants to infer AFS.  Secondly, the effectiveness of
imputation depends on the pattern of LD, which may lead to potential bias in
population genetical inferences. Thirdly, the current imputation algorithms are
slow. For a thousand samples, the fastest algorithm may be slower than read
mapping algorithms, which is frequently the bottleneck of analyzing NGS data
(Hyun Min Kang, personal communication). More samples and the use of more
accurate imputation algorithms will be even slower.

These potential concerns make us reconsider if imputation is always preferred.
We notice that we perform imputation mainly to reuse the methods developed for
genotyping data, but would it be possible to derive new methods to solve
classical medical and population genetical problems without precise genotypes?

Another application of NGS that requires genotype data is to discover somatic
mutations or germline mutations between a few related
samples~\citep{Ley:2008ve,Mardis:2009qf,Shah:2009cr,Pleasance:2010bh,Pleasance:2010dq,Roach:2010oq,Conrad:2011kx}.
For such an application, samples are often sequenced to high coverage.
Although it is not hard to achieve an error rate one per 100,000
bases~\citep{Bentley:2008cr}, mutations occur at a much lower rate, typically
of the order of $10^{-6}$ or even $10^{-7}$. Naively calling genotypes and then
comparing samples frequently would not work well~\citep{Ajay:2011fk}, because subtle uncertainty in
genotypes may lead to a bulk of errors. From another angle, however, when
discovering rare mutations, we only care about the difference between samples.
Genotypes are just a way of measuring the difference. Is it really necessary to
go through the genotype calling step?

This article explores the answer to these questions. We will show in the
following how to compute various statistics directly from sequencing data
without knowing genotypes. We will also evaluate our methods on real data.

\begin{methods}
\section{METHODS}
This section presents the precise equations on how to infer various statistics
such as the genotype frequency and AFS, and to perform various statistical test
such as testing HWE and associations. Some of these equations have already been
described in the existing literature, but for theoretical completeness, we give
the equations using our notations. The last subsection reviews the existing
methods and summarizes the differences between them, as well as between ours
and existing formulation.

In the Methods section, we suppose there are $n$ individuals with the $i$-th
individual having $m_i$ ploidy. At a site, the sequence data for the $i$-th
individual is represented as $d_i$ and the genotype is $g_i$ which is an
integer in $[0,m_i]$, equal to the number of reference alleles in the
individual.
Table~1 
gives notations common across this Methods section. The detailed derivation of
the equations in this article is presented in an online document
(http://bit.ly/stmath).

\begin{table}[!htb]\label{tab:notation}
\processtable{Common notations}
{\begin{tabular}{lp{7cm}}
\toprule
Symbol & Description \\
\midrule
$n$ & Number of samples \\
$m_i$ & Ploidy of the $i$-th sample ($1\le i\le n$)\\
$M$ & Total number of chromosomes in samples: $M=\sum_i m_i$\\
$d_i$ & Sequencing data (bases and qualities) for the $i$-th sample\\
$g_i$ & Genotype (the number of reference alleles) of the $i$-th sample \mbox{($0\le g_i\le m_i$)}$^1$\\
$\phi_k$ & Probability of observing $k$ reference alleles ($\sum_{k=0}^M\phi_k=1$) \\
$\Pr\{A\}$ & Probability of an event $A$\\
$\mathcal{L}_i(\theta)$ & Likelihood function for the $i$-th sample: $\mathcal{L}_i(\theta)=\Pr\{d_i|\theta\}$ \\
\botrule
\end{tabular}}{$^1$ In this article, we only consider biallelic variants.}
\end{table}

\subsection{Assumptions}

\subsubsection{Site independency} We assume data at different sites are
independent.  This may not be true in real data because sequencing and mapping
are context dependent; when there is an insertions or deletion (INDEL) error or
INDEL polymorphism, sites nearby are also correlated in alignment. Nonetheless,
most of the existing methods make this assumption for simplicity. The effect
of site dependency may also be reduced by post filtering and properly modeling
the mapping and alignment errors~\citep{Li:2008zr,Li:2011kx}.

\subsubsection{Error independency and sample independency}
We assume that at a site the sequencing and mapping errors of different reads
are independent. As a result the likelihood functions of different individuals
are independent:
\begin{equation}
\mathcal{L}(\theta)=\prod_{i=1}^n\mathcal{L}_i(\theta)
\end{equation}
In real data, errors may be dependent of sequence
context~\citep{Nakamura:2011kx}. The independency assumption may not hold. It
is possible to model error dependency within an individual~\citep{Li:2008zr},
but the sample independency assumption is essential to all the derivations
below.

\subsubsection{Biallelic variants}
We assume all variants are biallelic. In the human population, the fraction of
triallelic SNPs is about 0.2\%~\citep{Hodgkinson:2010uq}. The biallele assumption
does not have a big impact to the modeling of SNPs, though it may have a bigger
impact to the modeling of INDELs at microsatellites.

\subsection{Computing genotype likelihoods}

For one sample at a site, the sequencing data $d$ is composed of an array of
bases on sequencing reads plus their base qualities. As we only consider
biallelic variants, we may focus on the two most evident types of nucleotides
and drop the less evident types if present. Thus at any site we see at most two
types of nucleotides. This treatment is not optimal, but sufficient in
practice.

Suppose at a site there are $k$ reads. Without losing generality, let the first
$l$ bases ($l\le k$) be identical to the reference and the rest be different.
The error probability of the $j$-th read base is $\epsilon_j$.  Assuming error
independency, we can derive that
\begin{equation}\label{equ:glk}
\mathcal{L}(g)=\frac{1}{m^k}\prod_{j=1}^l\Big[(m-g)\epsilon_j+g(1-\epsilon_j)\Big]\prod_{j=l+1}^k\Big[(m-g)(1-\epsilon_j)+g\epsilon_j\Big]
\end{equation}
where $m$ is the ploidy.


\subsection{Inferences from multiple samples}

\subsubsection{Estimating the site allele frequency}
In this section we estimate the per-site reference allele frequency $\psi$.  For the
$i$-th sample, let $m_i$ be the ploidy, $g_i$ the genotype and $d_i$ the
sequencing data.  Assuming Hardy-Weinberg equilibrium (HWE), we can compute the
likelihood of $\psi$:
\begin{equation}\label{equ:flk}
\mathcal{L}(\psi)=\sum_{g_1}\cdots\sum_{g_n}\prod_i\Pr\{d_i,g_i|\psi\}=\prod_{i=1}^n\sum_{g=0}^{m_i}\mathcal{L}_i(g)f(g;m_i,\psi)
\end{equation}
where $\mathcal{L}_i(g_i)$ is computed by Eq.~\eqref{equ:glk} and
\begin{equation}
f(g;m,\psi)=\binom{m}{g}\psi^g(1-\psi)^{m-g}
\end{equation}
is the probability mass function of the binomial distribution ${\rm
Binom}(m,\psi)$.

Knowing the likelihood of $\psi$, we may numerically find the max-likelihood
estimate with, for example, Brent's method~\citep{Brent:1973kx}. An alternative
approach is to infer using an expectation-maximization algorithm (EM),
regarding the sample genotypes as missing data. Given we know the estimate
$\psi^{(t)}$ at the $t$-th iteration, the estimate at the $(t+1)$-th iteration
is
\begin{equation}\label{eq:saf-em}
\psi^{(t+1)}=\frac{1}{M}\sum_{i=1}^n\frac{\sum_{g}g\mathcal{L}_i(g)f(g;m_i,\psi^{(t)})}{\sum_{g}\mathcal{L}_i(g)f(g;m_i,\psi^{(t)})}
\end{equation}
where $M=\sum_im_i$ is the total number of chromosomes in samples.

When the signal from the data is strong, or equivalently for each $i$, one of
$\mathcal{L}_i(g)$ is much larger than others, the EM algorithm converges
faster than the direct numerical solution using Brent's method. However, when
the signal from the data is weak, numerical method may converge faster than
EM~\citep{Kim:2011fk}. In implementation, we apply 10 rounds of EM iterations.
If the estimate does not converge after 10 rounds, we switch to Brent's method.

\subsubsection{Estimating the genotype frequencies}
In this section, we assume all samples have the same ploidy: $m=m_1=\cdots=m_n$
and aim to estimate $\xi_g$, the frequency of genotype $g$. The likelihood of
$\{\xi_0,\ldots,\xi_m\}$ is:
\begin{equation}
\mathcal{L}(\xi_0,\ldots,\xi_m)=\prod_{i=1}^n\sum_{g=0}^{m}\mathcal{L}_i(g)\xi_g
\end{equation}
with the constraint $\sum_g\xi_g=1$. The EM iteration equation is
\begin{equation}\label{eq:sgf-em}
\xi^{(t+1)}_g=\frac{1}{n}\sum_{i=1}^n\frac{\mathcal{L}_i(g)\xi^{(t)}_g}{\sum_{g'}\mathcal{L}_i(g')\xi_{g'}^{(t)}}
\end{equation}

An important application of genotype frequencies is to test HWE
for diploid samples ($m=2$).  When genotypes are known, we can perform
a 1-degree $\chi^2$ test. This approach would not work for sequencing data as
it does not account for the uncertainty in genotypes, especially when the
average read depth of each individual is low. A
proper solution is to perform a likelihood-ratio test (LRT). The test statistic is
\begin{equation}\label{eq:hwe}
D_e=-2\log\frac{\mathcal{L}(\hat{\psi})}{\mathcal{L}(\hat{\xi_0},\hat{\xi_1},\hat{\xi_2})}
=-2\log\frac{\mathcal{L}((1-\hat{\psi})^2,2\hat{\psi}(1-\psi),\hat{\psi}^2)}{\mathcal{L}(\hat{\xi_0},\hat{\xi_1},\hat{\xi_2})}
\end{equation}
where
\begin{equation}\label{eq:psimax}
\hat{\psi}=\argmax_{\psi}\mathcal{L}(\psi)
\end{equation}
is the max-likelihood estimate of the site allele frequency and similarly
$\hat{\xi}_0$, $\hat{\xi}_1$ and $\hat{\xi}_2$ are the max-likelihood estimate
of the genotype frequencies. Because $\mathcal{L}(\hat{\psi})$ has one degree of freedom
and $\mathcal{L}(\hat{\xi_0},\hat{\xi_1},\hat{\xi_2})$ has two degree of freedom,
the $D_e$ statistic approximately follows the
1-degree $\chi^2$ distribution. For genotype data, $D_e$ approaches the standard
HWE test statistic computed from a 3-by-2 contingency table.

\subsubsection{Estimating haplotype frequencies between loci}
In this section, we assume all samples are diploid. Given $k$ loci, let
$\vec{h}=(h_1,\ldots,h_k)$ be a haplotype where $h_j$ equals 1 if the allele at
the $j$-th locus is identical to the reference, and equals 0 otherwise.  Let
$\eta_{\vec{h}}$ be the frequency of haplotype $\vec{h}$ satisfying
$\sum_{\vec{h}}\eta_{\vec{h}}=1$, where
$$
\sum_{\vec{h}}\eta_{\vec{h}}=\sum_{h_1=0}^1\sum_{h_2=0}^1\cdots\sum_{h_k=0}^1\eta_{(h_1,\ldots,h_k)}
$$
Knowing the genotype likelihood at the $j$-th locus for the $i$-th individual
$\mathcal{L}^{(j)}_i(g)$, we can compute the haplotype frequencies iteratively
with:
\begin{equation}\label{equ:hf}
\eta^{(t+1)}_{\vec{h}}=\frac{\eta_{\vec{h}}^{(t)}}{n}\sum_{i=1}^n\frac{\sum_{\vec{h}'}\eta_{\vec{h'}}^{(t)}\prod_{j=1}^k\mathcal{L}^{(j)}_i(h_j+h'_j)}
{\sum_{\vec{h}',\vec{h}''}\eta_{\vec{h'}}^{(t)}\eta_{\vec{h''}}^{(t)}\prod_{j}\mathcal{L}^{(j)}_i(h'_j+h''_j)}
\end{equation}
When sample genotypes are all certain, this EM iteration is reduced to the
standard EM for estimating haplotype frequencies using genotype
data~\citep{Excoffier:1995ly}.

The time complexity of computing Eq.~\eqref{equ:hf} is $O(n\cdot 4^k)$ and thus
it is impractical to estimate the haplotype frequency for many loci jointly.  A
typical use of Eq.~\eqref{equ:hf} is to measure linkage disequilibrium (LD)
between two loci.

\subsubsection{Testing associations}
Suppose we divide samples into two groups of size $n_1$ and $n-n_1$,
respectively, and want to test if group 1 significantly differs from group 2.
One possible test statistic could be~\citep{Kim:2010ve,Kim:2011fk}
\begin{equation}\label{eq:asso1}
D_{a1}=-2\log\frac{\mathcal{L}(\hat{\psi})}{\mathcal{L}^{[1]}(\hat{\psi}^{[1]})\mathcal{L}^{[2]}(\hat{\psi}^{[2]})}
\end{equation}
where $\hat{\psi}$ is the max-likelihood estimate of the site allele frequency
of all samples (Eq.~\ref{eq:psimax}), and $\hat{\psi}^{[1]}$ and
$\hat{\psi}^{[2]}$ are the estimates of allele frequency in group 1 and group
2, respectively.  Under the null hypothesis, $D$ approximately follows the
1-degree $\chi^2$ distribution.

A potential concern with the $D_{a1}$ statistic is that the computation of $\mathcal{L}(\psi)$
assumes HWE.  When HWE is violated, false positives may
arise~\citep{Nielsen:2011fk}. For diploid samples, a safer statistic is
\begin{equation}\label{eq:asso2}
D_{a2}=-2\log\frac{\mathcal{L}(\hat{\xi}_0,\hat{\xi_1},\hat{\xi}_2)}
{\mathcal{L}^{[1]}(\hat{\xi}_0^{[1]},\hat{\xi}_1^{[1]},\hat{\xi}_2^{[1]})\mathcal{L}^{[2]}(\hat{\xi}_0^{[2]},\hat{\xi}_1^{[2]},\hat{\xi}_2^{[2]})}
\end{equation}
which in principle follows the 2-degree $\chi^2$ distribution under the null
hypothesis.  However, when both cases and controls are in HWE, the degree of
freedom is reduced and this statistic is underpowered.

We have not found a powerful test statistic robust to HWE violation. For
practical applications, we propose to take the P-value computed with $D_{a1}$,
while filtering candidates having a low $D_{a2}$ to reduce false positives
caused by HWE violation (see Results).

\subsubsection{Estimating the number of non-reference alleles}
In this section we use the term \emph{site reference allele count} to refer to
the number of reference alleles at one single site. Allele count is a discrete number
while allele frequency is contiguous.

For convenience, define random vector $\vec{G}=(G_1,\ldots,G_n)$ to be a
genotype configuration, and $X=\sum_iG_i$ to be the site reference allele count
in all the samples.  Assuming HWE, we have
$$
\Pr\{\vec{G}=\vec{g}|X=k\}=\delta_{k,s_n(\vec{g})}\prod_{i=1}^n\frac{\binom{m_i}{g_i}}{\binom{M}{k}}
$$
where $s_n(\vec{g})=\sum_i g_i$ is the total number of reference alleles in a
genotype configuration $\vec{g}$, and $\delta_{kl}$ is the Kronecker delta
function which equals 1 if $k=l$ and equals 0 otherwise. The likelihood of
allele count is
\begin{equation}\label{equ:klk}
\mathcal{L}(k)=\Pr\{\vec{d}|X=k\}=\frac{1}{\binom{M}{k}}\sum_{g_1}\cdots\sum_{g_n}\delta_{k,s_n(\vec{g})}\prod_i\binom{m_i}{g_i}\mathcal{L}_i(g_i)
\end{equation}
where $\vec{d}=(d_1,\ldots,d_n)$ represents all sequencing data. To compute
this probability efficiently, we define
\begin{equation}\label{eq:z-def}
z_{jl}=\sum_{g_1=0}^{m_1}\cdots\sum_{g_j=0}^{m_j}\delta_{l,s_j(\vec{g})}\prod_{i=1}^j\binom{m_i}{g_i}\mathcal{L}_i(g_i)
\end{equation}
for $0\le l\le \sum_{i=1}^jm_i$ and $z_{jl}=0$ otherwise. $z_{jl}$ can be
calculated iteratively with
\begin{equation}\label{equ:z}
z_{jl}=\sum_{g_j=0}^{m_j}z_{j-1,l-g_j}\cdot\binom{m_j}{g_j}\mathcal{L}_j(g_j)
\end{equation}
starting from $z_{00}=1$. Comparing the definition of $z_{nk}$ and
Eq.~\eqref{equ:z}, we know that
\begin{equation}\label{equ:klk2}
\mathcal{L}(k)=\frac{z_{nk}}{\binom{M}{k}}
\end{equation}
which computes the likelihood of the allele count.

Although the computation of the likelihood function $\mathcal{L}(k)$ is more
complex than of $\mathcal{L}(\psi)$, $\mathcal{L}(k)$ is discrete, which is
more convenient to maximize or sum over. This likelihood function establishes
the foundation of the Bayesian inference.

\subsubsection{Numerical stability of the allele count estimation}
When computing $z_{jl}$ with Eq.~\eqref{equ:z}, floating point underflow may
occur given large $j$. A numerically stable approach is to compute
$y_{jl}=z_{jl}/\binom{M_j}{l}$ instead, where $M_j=\sum_{i=1}^j m_i$. Thus
\begin{equation}\label{equ:klky}
\mathcal{L}(k)=y_{nk}
\end{equation}
and by replacing $z_{jk}$ with $y_{jk}\binom{M_j}{l}$ in Eq.~\eqref{equ:z}, we can derive:
\begin{eqnarray}\label{equ:y}
y_{jk}&=&\left(\prod_{l=0}^{m_j-1}\frac{k-l}{M_{j}-l}\right)\sum_{g_j=0}^{m_j}y_{j-1,k-g_j}\cdot\binom{m_j}{g_j}\mathcal{L}_j(g_j)\\\nonumber
	&&\cdot\left(\prod_{l=g_j}^{m_j-1}\frac{M_{j-1}-k+l+1}{k-l}\right)
\end{eqnarray}
However, we note that $y_{jl}$ may decrease exponentially with increasing $j$.
Floating point underflow may still occur. An even better solution is to rescale
$y_{jl}$ for each $j$, similar to the treatment of the forward algorithm for
Hidden Markov Models~\citep{Durbin:1998uq}. In practical implementation, we
compute
\begin{equation}\label{eq:rescale}
\tilde{y}_{jl}=\frac{y_{jl}}{\prod_{j'=1}^jt_{j'}}
\end{equation}
where $t_{j}$ is chosen such that $\sum_{l}\tilde{y}_{jl}=1$.

As another implementation note, most $y_{jl}$ are close to zero and thus
$y_{nk}$ can be computed in a band rather than in a triangle. This may
dramatically speed up the computation of the likelihood.

\subsubsection{Calling variants}
In variant calling, we have a strong prior knowledge that at most of sites all
samples are homozygous to the reference.  To utilize the prior knowledge, we
may adopt a Bayesian inference for variant calling.  Let $\phi_k$, $k=1,\ldots,M$,
be the probability of seeing $k$ reference alleles among $M$
chromosomes/haplotypes. For convenience, define $\Phi=\{\phi_k\}$, which is in
fact the sample \emph{allele frequency spectrum} (AFS) for $M$ chromosomes.
Recall that $X$ is the number of reference alleles in the samples. The posterior of
$X$ is
\begin{equation}\label{eq:post}
\Pr\{X=k|\vec{d},\Phi\}=\frac{\phi_k\Pr\{\vec{d}|X=k\}}{\sum_l\phi_l\Pr\{\vec{d}|X=l\}}
=\frac{\phi_k\mathcal{L}(k)}{\sum_l\phi_l\mathcal{L}(l)}
\end{equation}
where $\mathcal{L}(k)$ is defined by Eq.~\eqref{equ:klk} and computed by
Eq.~\eqref{equ:klky}.  In variant calling, we define \emph{variant quality} as
$$
Q_{\rm var}=-10\log_{10}\Pr\{X=M|\vec{d},\Phi\}
$$
and call the site as a variant if $Q_{\rm var}$ is large enough. Because in
deriving $\mathcal{L}(k)$, we do not require the ploidy of each sample to be
the same. The variant calling method described here are in theory applicable to
pooled resequencing with unequal pool sizes.

\subsubsection{Estimating the sample allele frequency spectrum (AFS)}
For variant calling (Eq.~\ref{eq:post}), we typically take the Wright-Fisher AFS as the prior. We
can also estimate the sample AFS with the maximum-likelihood inference when the
Wright-Fisher prior deviates from the data.

Suppose we have $L$ sites of interest and we want to estimate the frequency
spectrum across these sites.  Let $X_a$, $a=1,\ldots,L$, be a random variable
representing the number of reference alleles at site $a$. We can use an EM
algorithm to find $\Phi$ that maximizes
$\Pr\{{\bf d}|\Phi\}$, the probability of data across all samples and all sites conditional on AFS.
The iteration equation is
\begin{equation}\label{eq:afs}
\phi_k^{(t+1)}=\frac{1}{L}\sum_a\Pr\{X_a=k|{\bf d},\Phi^{(t)}\}
\end{equation}
We call this method of estimating AFS as \emph{EM-AFS}. Alternatively, we may
also acquire the max-likelihood estimate of the allele count at each site using
Eq.~\eqref{equ:klk2}. The normalized histogram of these counts gives the AFS.
We call this method as \emph{site-AFS}. We will compare the two methods in the
Results section.


\subsection{Discovering somatic and germline mutations}
One of the key goals in cancer resequencing is to identify the somatic
mutations between a normal-tumor sample pair~\citep{Robison:2010ys}, which can
be achieved by computing a likelihood ratio.  Given a pair of samples, 
the following likelihood ratio is an informative score:
\begin{equation}\label{eq:dp}
D_p=-2\log\frac{\mathcal{L}^{[1]}(\hat{g})\mathcal{L}^{[2]}(\hat{g})}{\mathcal{L}^{[1]}(\hat{g}^{[1]})\mathcal{L}^{[2]}(\hat{g}^{[2]})}
\end{equation}
where $\mathcal{L}^{[\cdot]}(g)$ is computed by Eq.~\eqref{equ:glk}, $\hat{g}$
maximizes $\mathcal{L}^{[1]}(g)\mathcal{L}^{[2]}(g)$, and similarly
$\hat{g}^{[1]}$ and $\hat{g}^{[2]}$ maximize $\mathcal{L}^{[1]}(g)$ and
$\mathcal{L}^{[2]}(g)$, respectively.

Note that in most practical cases, $\hat{g}$ equals either
$\hat{g}^{[1]}$ or $\hat{g}^{[2]}$.  When this stands, we have:
$$
\mathcal{L}^{[1]}(\hat{g})\mathcal{L}^{[2]}(\hat{g})=\max\big\{
\mathcal{L}^{[1]}(\hat{g}^{[1]})\mathcal{L}^{[2]}(\hat{g}^{[1]}), \mathcal{L}^{[1]}(\hat{g}^{[2]})\mathcal{L}^{[2]}(\hat{g}^{[2]})
\big\}
$$
and then we can prove:
$$
D_p=2\log\left\{ \min\left\{\frac{\mathcal{L}^{[1]}(\hat{g}^{[1]})}{\mathcal{L}^{[1]}(\hat{g}^{[2]})},\frac{\mathcal{L}^{[2]}(\hat{g}^{[2]})}{\mathcal{L}^{[2]}(\hat{g}^{[1]})}\right\}\right\}
$$
This equation has an intuitive interpretation: we are certain about a candidate
somatic mutation only if both genotypes in both samples are clearly better
than other possible genotypes.

A natural extension to discovering somatic mutations is to discover {\it de
novo} and somatic mutations in a family trio~\citep{Conrad:2011kx}.  To
identify such mutations, we may compute the maximum likelihoods of genotype
configurations without the family constraint and with the constraint, and then
take the ratio between the two resulting likelihoods. The larger the ratio, the
more confident the mutation. More exactly, the likelihood ratio is:
\begin{equation}\label{eq:dt}
D_t=-2\log\frac
{\max_{(g_c,g_f,g_m)\in\mathcal{G}}\{\mathcal{L}_c(g_c)\mathcal{L}_f(g_f)\mathcal{L}_m(g_m)\}}
{\max\mathcal{L}_c(g_c)\cdot\max\mathcal{L}_f(g_f)\cdot\max\mathcal{L}_m(g_m)}
\end{equation}
where $\mathcal{L}_c(g_c)$, $\mathcal{L}_f(g_f)$ and $\mathcal{L}_m(g_m)$
are the child, father and mother genotype likelihoods respectively, and
$\mathcal{G}$ is the set of genotype configurations satisfying the Mendelian
inheritance.

Although most of the derivation in this article assumes variants are biallelic,
we drop this assumption in the implementation for methods described in this subsection.  We have
observed false somatic/germline mutations caused by the mismodeling of
triallelic variants (Mark Depristo, personal communication). The biallelic
assumption may lead to false positives.

\subsection{Working with diploid multi-allelic sites}
Suppose at a site there are $p$ alleles. The site frequency of allele $h$ being $\psi_h$ with
$\sum_h\psi_h=1$. If we assume the site is under the Hardy-Weinberg equilibrium, the likelihood function of $\{\psi_1,\ldots,\psi_p\}$ is:
\begin{equation}
\mathcal{L}(\psi_1,\ldots,\psi_p)=\prod_{i=1}^n\sum_{h,h'}\mathcal{L}_i(\langle h,h'\rangle)\psi_h\psi_{h'}
\end{equation}
where $\langle h,h'\rangle$ represents a pair of unordered integers, or a
diploid genotype.  The EM iteration equation can be derived as:
\begin{equation}
\psi^{(t+1)}_h=\frac{1}{n}\sum_{i=1}^n\frac{\sum_{h'}\mathcal{L}(\langle h,h'\rangle)\psi^{(t)}_h\psi^{(t)}_{h'}}{\sum_{k,k'}\mathcal{L}(\langle k,k'\rangle)\psi^{(t)}_k\psi^{(t)}_{k'}}
\end{equation}
To test whether a site is multi-allelic, we may compute the likelihood ratio
\begin{equation}
D_{m2}=-2\log\frac{\mathcal{L}(\psi_1,\psi_2)}{\mathcal{L}(\psi_1,\psi_2,\psi_3)}
\end{equation}
which approximately follows a 1-degree $\chi^2$ distribution. Similarly,
we may perform variant calling by testing
\begin{equation}
D_{m1}=-2\log\frac{\mathcal{L}(0,1)}{\mathcal{L}(\psi_1,\psi_2)}
\end{equation}
as is proposed by \citet{Kim:2010ve}.

\subsection{Related works}
During SNP calling, Thunder~\citep{Li:2011fk} and glfMultiples
(http://bit.ly/glfmulti) compute the site allele frequency by numerically
maximizing the likelihood (Eq.~\ref{equ:glk}). Genome Analysis Toolkit (GATK;
\citealp{Depristo:2011vn}) infers the frequency with EM (Eq.~\ref{eq:saf-em}).
\citet{Kim:2011fk} infers the frequency with both the numerical and the EM
algorithms.  \citet{Li:2010ys} derived an alternative method to estimate the
site allele frequency, which is not covered in this article.
SeqEM~\citep{Martin:2010dz} estimates the genotype frequency using EM
(Eq.~\ref{eq:sgf-em}) with a different parameterization.  \citet{Le:2010uq}
derived Eq.~\eqref{equ:klk2}. The conclusion is correct, but the derivation is
not rigorous: the binomial coefficient in Eq.~\eqref{equ:klk} was left out.
\citet{Yi:2010zr} came to a similar set of equations to Eq.~\eqref{equ:z} and
\eqref{eq:post}, but the prior is taken from the estimated site allele
frequency.  To the best of our knowledge, \citet{Kim:2010ve} is the first to
use genotype likelihood based LRT to compute P-value of associations
(Eq.~\ref{eq:asso1}) with more thorough evaluation in a recent
paper~\citep{Kim:2011fk}. \citet{Nielsen:2011fk} further proposed to test
associations with a score test~\citep{Schaid:2002qf}.  Except
\citet{Kim:2010ve}, all the previous works focus on diploid samples, while many
equations in this article can be in theory applied to multi-ploidy samples and
pooled samples.

In this article, our contribution includes testing HWE, estimating haplotype
frequency, the proposal of two-degree association test, a simple but effective
model for discovering somatic mutations, the rigorous derivation and
numerically stable implementation of a discrete allele count estimator, and an
EM algorithm for inferring AFS.

\end{methods}

\section{RESULTS}
\subsection{Implementation}
Most of equations for diploid samples ($m=2$) have been implemented in the
SAMtools software package~\citep{Li:2009ys}, which is distributed under the MIT
open source license, free to both academic and commercial uses. The exact
Eq.~\eqref{equ:klky}--\eqref{eq:rescale} have also been implemented in GATK as
the default SNP calling model.

The SAMtools package consists of two key components {\tt samtools} and {\tt
bcftools}.  The former computes the genotype likelihood $\mathcal{L}(g)$ using
an improved version of Eq.~\eqref{equ:glk} which considers error dependencies;
the latter component calls variants and infers various statistics described in this
article. To clearly separate the two steps, we designed a new \emph{Binary
variant call format} (BCF), which is the binary representation of the variant
call format (VCF;~\citealp{Danecek:2011fk}) and is more compact and much faster to
process than VCF.  On real data, computing genotype likelihoods especially for
INDELs is typically 10 times slower than variant calling. The separation of
genotype likelihood computation and subsequent inferences enhances the
flexibility and improves the efficiency for inferring AFS. {\tt Bcftools} also
directly works with VCF files, but is less efficient than with BCF files.

Table 2 shows how VCF information tags generated by SAMtools are related
to the equations in this article. We refer to the SAMtools manual page
for detailed description.

\begin{table}[!htb]\label{tab:notation}
\processtable{SAMtools specific VCF information}
{\begin{tabular}{llp{5.6cm}}
\toprule
INFO$^1$ & Equation$^2$ & Description \\
\midrule
{\tt AF1} & \ref{equ:flk},\ref{eq:saf-em} & Non-reference site allele frequency \\
{\tt G3} & \ref{eq:sgf-em} & Diploid genotype frequency \\
{\tt HWE} & \ref{eq:hwe} & P-value of Hardy-Weinberg equilibrium \\
{\tt NEIR} & \ref{equ:hf} & Neighboring $r^2$ linkage disequilibrium statistic \\
{\tt LRT} & \ref{eq:asso1} & 1-degree association test P-value \\
{\tt LRT2} & \ref{eq:asso2} & 2-degree association test P-value \\
{\tt AC1} & \ref{equ:klky},\ref{equ:y},\ref{eq:rescale} & Non-reference site allele count \\
{\tt FQ} & \ref{eq:post} & Prob. of the site being poly. among samples \\
{\tt CLR} & \ref{eq:dp},\ref{eq:dt} & Log likelihood ratio score for {\it de novo} mutations \\
\botrule
\end{tabular}}{
$^1$ Tag at the VCF additional information field (INFO)\\
$^2$ Related, though not exact, equations for computing the values
}
\end{table}

\subsection{Inferring the allele count}

\begin{figure}[!htb]
\centering
\includegraphics[width=.40\textwidth]{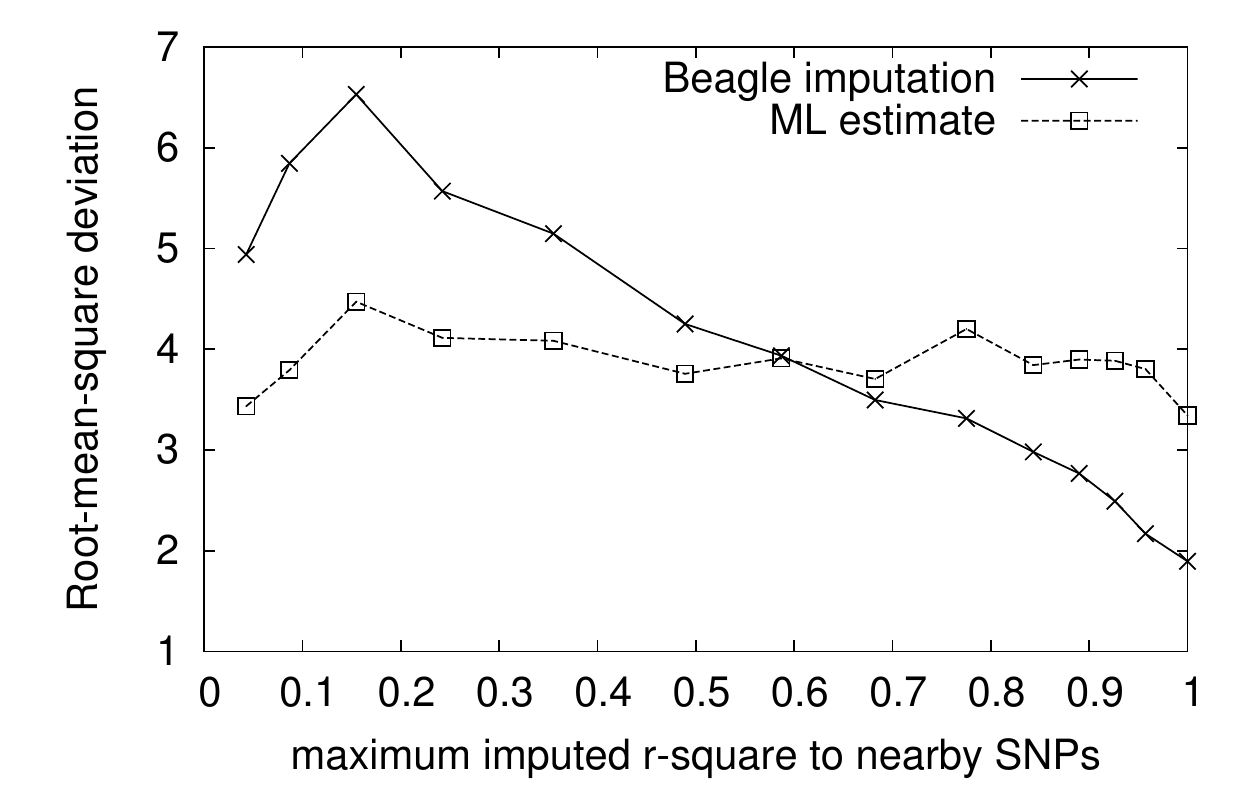}
\caption{Correlation of the site allele count accuracy with LD. The site allele
count is estimated with Beagle imputation (solid line) and with
Eq.~\eqref{equ:klk2} (dashed line) at sites typed by the Omni genotyping chip.
For each Omni SNP, the maximum $r^2$ LD statistic between the SNP and
20 nearby SNPs called by SAMtools (10 upstream and 10 downstream) is computed from imputed genotypes.
Omni SNPs are then ordered by the maximum $r^2$ and approximately evenly
divided into 15 bins.  For each bin, the root-mean-square deviation between the
Omni allele count and the estimated allele count is computed as a measurement
of the allele count accuracy.}\label{fig:ac}
\end{figure}

We downloaded the chromosome 20 alignments of 49 Pilot-1 CEU samples sequenced
by the 1000 Genomes Project using the Illumina technology only. We called the
SNPs with SAMtools and imputed the genotypes with Beagle under the default
settings. At 32,522 sites genotyped using the Omni genotyping chip and
polymorphic in the 49 samples, the root-mean-square deviation (RMSD) between
the allele count acquired from Omni genotypes and the estimate using
Eq.~\eqref{equ:klk2} equals 3.7, the same as the RMSD between the Omni and the
Beagle-imputed genotypes.  Not surprisingly, imputed genotypes are more
accurate when there is a tightly linked SNP nearby, while the imputation-free
estimate is less affected (Fig.~\ref{fig:ac}).

However, on the unreleased European data from the 1000 Genomes Project
consisting of 670 samples, Beagle imputation is better than our imputation-free
method (RMSD(imput)=12.7; RMSD(imput-free)=15.0).  We conjure that this is
because with more samples, it is more frequent for two samples to share
a long haplotype.  The LD plays a more important role in counteracting the lack
of coverage.  Nonetheless, we should beware that sites selected on the Omni
genotyping chip may not be a good representative of all SNPs. For example, for
the sites on the Omni chip, only 8\% of SNPs do not have a nearby SNP with
$r^2>0.05$ in a 20-SNP window (the `nearby SNPs' include all SNPs discovered in
the 670 samples), but this percentage is increased to 30\% for all SNPs.  The
large fraction of unlinked SNPs might hurt the accuracy of imputation based
methods.

We have also evaluated our method on an unpublished target reqsequencing data
set consisting of about 2000 samples~(Haiman et al., personal communication).
The imputation based method does not perform well (RMSD(imput)=54.8;
RMSD(imput-free)=42.5), probably due to the lack of linked SNPs around
fragmented target regions.

\subsection{Inferring the allele frequency spectrum}
\begin{figure}[tb]
\centering
\includegraphics[width=.40\textwidth]{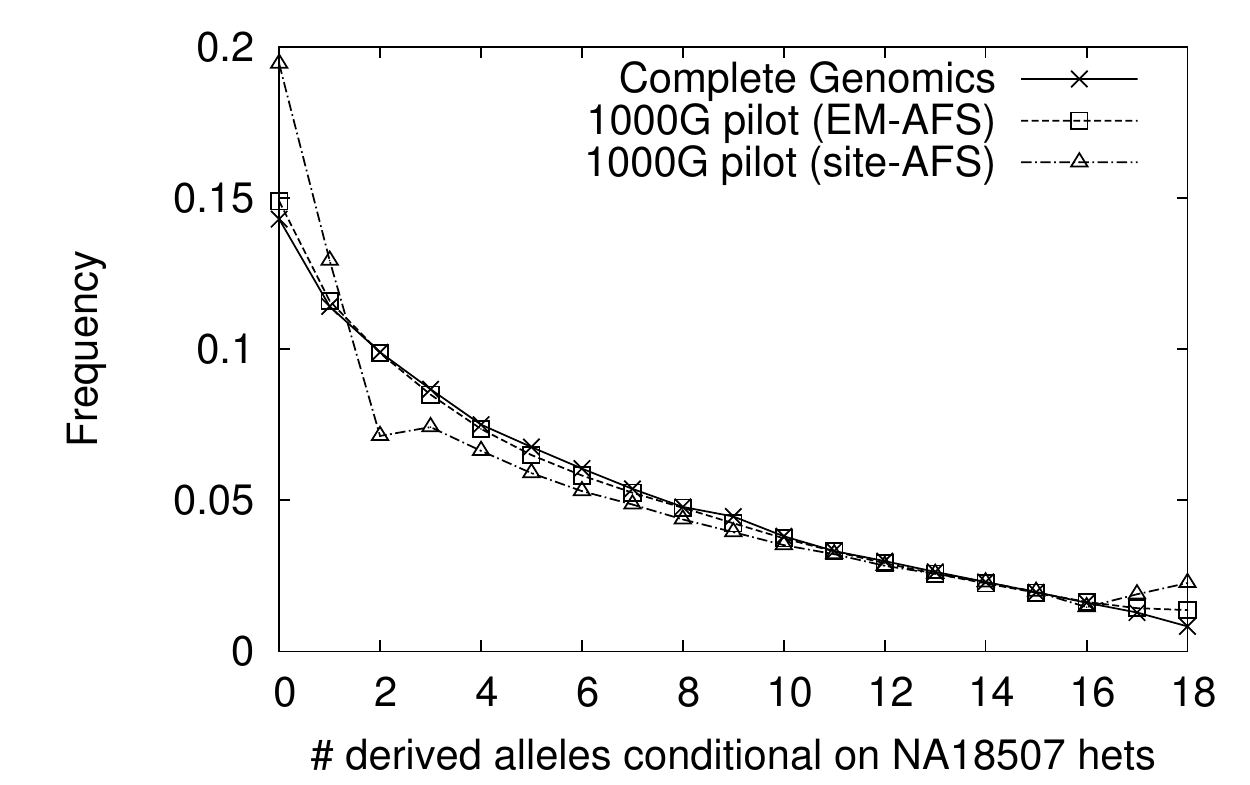}
\caption{The derived AFS conditional on heterozygotes discovered in the NA18507
genome~(\citealp{Bentley:2008cr}; AC:SRA000271). Heterozygotes were called with SAMtools on
BWA~\citep{Li:2009uq} alignment.  The ancestral sequences were
determined from the Ensembl EPO alignment~\citep{Paten:2008uq}, with the requirement of the
chimpanzee and orangutan sequences being identical.  The AFS at these heterozygotes were
computed in three ways: a) from the 9 independent Yoruba individuals sequenced by Complete
Genomics~\citep{Drmanac:2010ly} and analyzed using CGA Tools version 1.10.0; b)
from 9 random Pilot-1 Yoruba individuals released by the 1000 Genomes Project
using the EM-AFS method and c) from the same 9 Pilot-1 individuals using
site-AFS.}\label{fig:afs}
\end{figure}

To evaluate the accuracy of the estimated allele frequency spectrum (AFS), we
compared the AFS obtained from the low-coverage data produced by the 1000
Genomes Project and from the high-coverage data released by Complete Genomics (http://bit.ly/m7LzvF).
Fig.~\ref{fig:afs} reveals that we can infer a fairly accurate AFS using the
EM-AFS method with 3-fold coverage per sample. On the other hand, the site-AFS
estimate is less stable, though the overall trend looks right. To estimate
properties across multiple sites, summing over the posterior distribution using EM-AFS is
more appropriate.

\subsection{Performing association test}
\begin{figure}[tb]
\centering
\includegraphics[width=.47\textwidth]{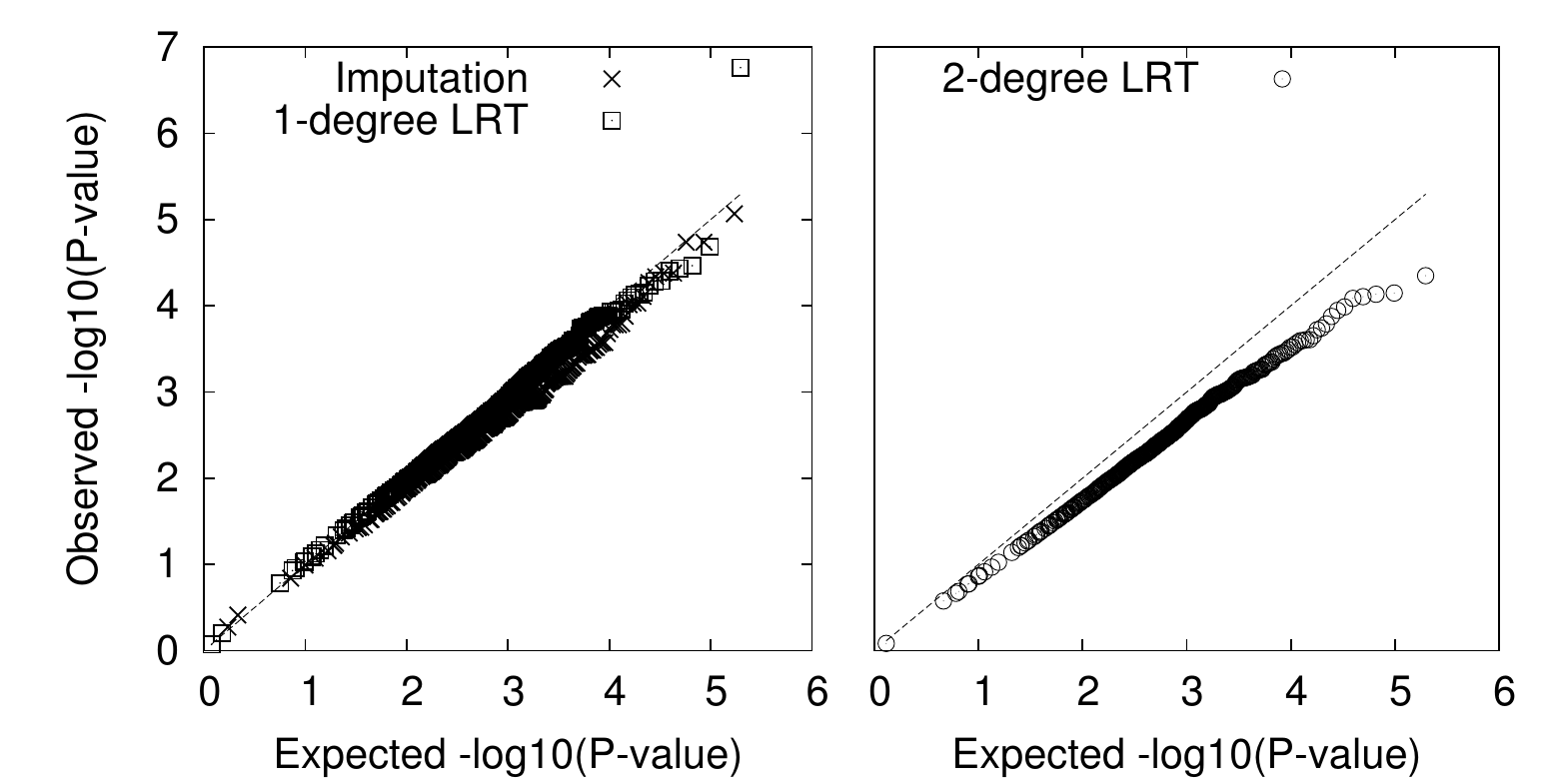}
\caption{QQ-plot comparing the association test statistics to the one-degree
and the two-degree $\chi^2$ distribution.
49 CEU samples sequenced by the 1000 Genomes Project using the Illumina
technology were randomly assigned to two groups of size 24 and 25,
respectively.  In the left, two association test statistics were computed on chromosome 20
between the two groups: one by the 1-degree likelihood ratio test (Eq.~\ref{eq:asso1}) and the other
by the canonical 1-degree $\chi^2$ test based on Beagle imputed
genotypes; in the right, the 2-degree likelihood rate test statistic (Eq.~\ref{eq:asso2}).}\label{fig:at}
\end{figure}

To evaluate the performance of the association test statistics $D_{a1}$
(Eq.~\ref{eq:asso1}), we constructed a perfect negative control using the
1000 Genomes data and derived the empirical distribution of $D_{a1}$.
We expect to see no associations.
Fig.~\ref{fig:at} shows that $D_{a1}$ largely follows the 1-degree $\chi^2$
distribution.  However, this method also produces one false positive SNP
($P<10^{-6}$). Closer investigation reveals that the SNP significantly violates
HWE ($P<10^{-6}$, computed with Eq.~\eqref{eq:hwe}), and thus violates the
assumption behind the derivation of $D_{a1}$.  In fact, if we test the
association with $D_{a2}$ which does not assume HWE, the false positive will be
suppressed ($P>0.001$). To test association using the 1-degree likelihood-ratio
test statistic, it is important to control HWE.

\subsection{Comparing sequencing data from the same individual}

\subsubsection{Comparing data sets of similar characteristics}
We acquired the NA12878 data used by~\citet{Depristo:2011vn}. This sample was sequenced
with HiSeq2000 using two libraries with each put on 8 lanes and each sequenced
to about 30-fold coverage.  We split the data in two by library and computed
$D_p$ (Eq.~\ref{eq:dp}) at each base on chromosome 20 to identify sites that
are called differently between the two libraries.  With a stringent threshold
$D_p\ge 30$ and without any filtering, 32 differences are called between
the libraries and most from the centromere. Because the libraries were made
from the same DNA at almost the same time, we expect to see no difference
between the libraries. Seeing 32 differences is very unlikely. To explore if
this is due to mismapping, we extracted reads around the 32 sites and remapped
them with BWA-SW~\citep{Li:2010fk}. 4 differences remain around the centromere,
implying that most of the differences between libraries are caused by the
variation in read mapping.  We further mapped the reads around the 4 sites to a
version of the human reference genome used by the 1000 Genomes Project for
phase-2 mapping (http://bit.ly/GRCh37d).  No differences
are left. This exercise reveals that when we come to very rare events,
mapping errors, instead of sequencing errors, lead to most of the artifacts.

\subsubsection{Comparing data sets of different characteristics}
We also did a harder version of the exercise above: comparing this 60-fold
HiSeq data to the old Illumina data for the same individual obtained more than
two years ago by the 1000 Genomes Project. We note that although DNA used in
the two data sets was originated from the same individual, somatic mutations in
cell lines, which is of the order of 1,000 per diploid
genome~\citep{Conrad:2011kx}, may be present. If the cell lines used in two
studies have greatly diverged, we might see up to a dozen somatic mutations
on chromosome 20.

This time with a threshold $D_p\ge 30$ and a maximum depth filter $150$, we
identified 667 single-base differences between the two data sets, far more than
our expectation. Again we sought to reduce mapping errors by remapping reads
with BWA-SW to the 1000 Genomes Project phase-2 reference genome. The number of
differences between the HiSeq and the old Illumina data quickly drops to 33.
If we further filter out clustered SNPs using a 100bp window, 13 potential differences
are left, 2\% of the initial candidates. This exercise again proves that
mismapping is the leading source of errors.

To see if the simple likelihood ratio (Eq.~\ref{eq:dp}) is comparable to more
sophisticated methods, we briefly tried SomaticSniper~\citep{Larson:2011xx} on
our data.  With a somatic score cutoff $65$, which is about $30$ in the
`$2\log$' scale as in $D_p$, SomaticSniper identified 1,826 differences.
SAMtools called fewer because it limits the mapping quality of reads with
excessive mismatches and applies base alignment quality~\citep{Li:2011kx} to
fix alignment errors around INDELs. With the two features switched off,
SAMtools called 1,696 differences, half of which overlap the differences found
by SomaticSniper. Calls unique to one method tend to have a mutation score close
to the threshold.

\section{DISCUSSIONS}

We have proposed a statistical framework for SNP calling as well as analyzing
sequencing data but without explicitly calling SNPs or their genotypes. With
this framework, we can discover somatic and germline mutations with appropriate
input data, efficiently estimate site allele frequency, allele
frequency spectrum and linkage disequilibrium, and test Hardy-Weinberg
equilibrium and association. On real data, we have demonstrated that our method
is able to achieve comparable accuracy to the best alternative methods.  We
have also extensively evaluated the performance of our method on several
unpublished data sets and got sensible results. Thus we conclude that useful
information can be obtained directly from sequencing data without SNP calling
or imputation.

Here we also want to emphasize a few findings in our evaluation of the methods.
Firstly, we confirmed that imputation is a viable method for transferring our
knowledges on genotyping data to low-coverage sequencing data.  It is likely to
have higher accuracy than our method given homogeneous whole-genome data
consisting of many samples. Nonetheless, we showed that the accuracy of
imputation depends on the LD nearby, which has long been speculated but without
direct evidence from real data until our work. Secondly, our proposed EM-AFS
method is able to accurately estimate AFS from low-coverage sequencing data. It
is more appropriate than estimating the site frequency separately and then
doing a histogram.  Thirdly, we observed that violation of HWE may cause false
positives in association mapping with the one-degree likelihood ratio
test~\citep{Kim:2011fk}. A two-degree likelihood ratio test is a conservative
way to avoid such an artifact. At last, we highlighted the importance of using
data of similar characteristics in the discovery of somatic mutations. We also
want to put a particular emphasis on the necessity of controlling mapping
errors when looking for very rare events such as somatic mutations, germline
mutations and RNA editing. It may be necessary to use two distinct mapping
algorithms to call variants and then take the intersection.

Frequently we require to know the exact DNA sequences or genotypes only to
estimate parameters or compute statistics. In these cases, the sequences and
genotypes are just intermediate results. When the sequence itself is uncertain,
mostly due to the uncertainty in sequencing and mapping, it may sometimes be
preferred to directly work with the uncertain sequence which may carry more
information than an arbitrarily ascertained sequence. We have showed that many
population genetical parameters and statistical tests can be adapted to work on
uncertain sequences, and believe more existing methods can be adapted in a
similar manner. Knowing the exact sequence is convenient, but not always
indispensable.

\section*{ACKNOWLEDGEMENTS}
We are grateful to Christopher Haiman for providing the unpublished data set
for assessing the performance, to Petr Danecek for evaluating the methods in
this article on large-scale data sets, to Rasmus Nielsen for the observation on
the occasional slow convergence of the EM algorithm, and to Si Quang Le and
Richard Durbin for the help on understanding the QCall model.  We also thank
the 1000 Genomes Project analysis subgroup and the GSA team at Broad Institute
for various helpful discussions, and thank all the SAMtools users for
evaluating the software package.

\paragraph{Funding\textcolon} NIH 1U01HG005208-01.
\bibliography{samtools}

\begin{thebibliography}{}

\bibitem[{1000 Genomes Project Consortium},
  2010]{1000-Genomes-Project-Consortium:2010qc}
{1000 Genomes Project Consortium} (2010).
\newblock A map of human genome variation from population-scale sequencing.
\newblock {\em Nature}, 467:1061--73.

\bibitem[Ajay et~al., 2011]{Ajay:2011fk}
Ajay, S.~S. et~al. (2011).
\newblock Accurate and comprehensive sequencing of personal genomes.
\newblock {\em Genome Res}.

\bibitem[Bentley et~al., 2008]{Bentley:2008cr}
Bentley, D.~R. et~al. (2008).
\newblock Accurate whole human genome sequencing using reversible terminator
  chemistry.
\newblock {\em Nature}, 456:53--9.

\bibitem[Brent, 1973]{Brent:1973kx}
Brent, R.~P. (1973).
\newblock {\em Algorithms for Minimization without Derivatives}.
\newblock Prentice-Hall, Englewood Cliffs, New Jersey.

\bibitem[Browning and Yu, 2009]{Browning:2009jl}
Browning, B.~L. and Yu, Z. (2009).
\newblock Simultaneous genotype calling and haplotype phasing improves genotype
  accuracy and reduces false-positive associations for genome-wide association
  studies.
\newblock {\em Am J Hum Genet}, 85:847--61.

\bibitem[Conrad et~al., 2011]{Conrad:2011kx}
Conrad et~al. (2011).
\newblock Variation in genome-wide mutation rates within and between human
  families.
\newblock {\em Nat Genet}, 43:712--714.

\bibitem[Danecek et~al., 2011]{Danecek:2011fk}
Danecek, P. et~al. (2011).
\newblock The variant call format and vcftools.
\newblock {\em Bioinformatics}.

\bibitem[Depristo et~al., 2011]{Depristo:2011vn}
Depristo, M.~A. et~al. (2011).
\newblock A framework for variation discovery and genotyping using
  next-generation {DNA} sequencing data.
\newblock {\em Nat Genet}, 43:491--8.

\bibitem[Drmanac et~al., 2010]{Drmanac:2010ly}
Drmanac, R. et~al. (2010).
\newblock Human genome sequencing using unchained base reads on self-assembling
  {DNA} nanoarrays.
\newblock {\em Science}, 327:78--81.

\bibitem[Durbin et~al., 1998]{Durbin:1998uq}
Durbin, R. et~al. (1998).
\newblock {\em Biological sequence analysis}.
\newblock Cambridge University Press.

\bibitem[Excoffier and Slatkin, 1995]{Excoffier:1995ly}
Excoffier, L. and Slatkin, M. (1995).
\newblock Maximum-likelihood estimation of molecular haplotype frequencies in a
  diploid population.
\newblock {\em Mol Biol Evol}, 12:921--7.

\bibitem[Hodgkinson and Eyre-Walker, 2010]{Hodgkinson:2010uq}
Hodgkinson, A. and Eyre-Walker, A. (2010).
\newblock Human triallelic sites: evidence for a new mutational mechanism?
\newblock {\em Genetics}, 184:233--41.

\bibitem[Howie et~al., 2009]{Howie:2009mb}
Howie, B.~N. et~al. (2009).
\newblock A flexible and accurate genotype imputation method for the next
  generation of genome-wide association studies.
\newblock {\em PLoS Genet}, 5:e1000529.

\bibitem[Kim et~al., 2010]{Kim:2010ve}
Kim, S.~Y. et~al. (2010).
\newblock Design of association studies with pooled or un-pooled
  next-generation sequencing data.
\newblock {\em Genet Epidemiol}, 34:479--91.

\bibitem[Kim et~al., 2011]{Kim:2011fk}
Kim, S.~Y. et~al. (2011).
\newblock Estimation of allele frequency and association mapping using
  next-generation sequencing data.
\newblock {\em BMC Bioinformatics}, 12:231.

\bibitem[Larson et~al., 2011]{Larson:2011xx}
Larson, D. et~al. (2011).
\newblock {SomaticSniper}: Identification of somatic point mutations in whole
  genome sequencing data.
\newblock {\em Submitted}.

\bibitem[Le and Durbin, 2010]{Le:2010uq}
Le, S.~Q. and Durbin, R. (2010).
\newblock {SNP} detection and genotyping from low-coverage sequencing data on
  multiple diploid samples.
\newblock {\em Genome Res}.

\bibitem[Ley et~al., 2008]{Ley:2008ve}
Ley, T.~J. et~al. (2008).
\newblock {DNA} sequencing of a cytogenetically normal acute myeloid leukaemia
  genome.
\newblock {\em Nature}, 456:66--72.

\bibitem[Li, 2011]{Li:2011kx}
Li, H. (2011).
\newblock Improving {SNP} discovery by base alignment quality.
\newblock {\em Bioinformatics}, 27:1157--8.

\bibitem[Li and Durbin, 2009]{Li:2009uq}
Li, H. and Durbin, R. (2009).
\newblock Fast and accurate short read alignment with burrows-wheeler
  transform.
\newblock {\em Bioinformatics}, 25:1754--60.

\bibitem[Li and Durbin, 2010]{Li:2010fk}
Li, H. and Durbin, R. (2010).
\newblock Fast and accurate long-read alignment with burrows-wheeler transform.
\newblock {\em Bioinformatics}, 26:589--95.

\bibitem[Li et~al., 2008]{Li:2008zr}
Li, H. et~al. (2008).
\newblock {Mapping short DNA sequencing reads and calling variants using
  mapping quality scores}.
\newblock {\em Genome Res}, 18:1851--8.

\bibitem[Li et~al., 2009a]{Li:2009ys}
Li, H. et~al. (2009a).
\newblock The sequence alignment/map format and samtools.
\newblock {\em Bioinformatics}, 25:2078--9.

\bibitem[Li et~al., 2009b]{Li:2009gb}
Li, Y. et~al. (2009b).
\newblock Genotype imputation.
\newblock {\em Annu Rev Genomics Hum Genet}, 10:387--406.

\bibitem[Li et~al., 2010a]{Li:2010ky}
Li, Y. et~al. (2010a).
\newblock {MaCH}: using sequence and genotype data to estimate haplotypes and
  unobserved genotypes.
\newblock {\em Genet Epidemiol}, 34:816--34.

\bibitem[Li et~al., 2010b]{Li:2010ys}
Li, Y. et~al. (2010b).
\newblock Resequencing of 200 human exomes identifies an excess of
  low-frequency non-synonymous coding variants.
\newblock {\em Nat Genet}, 42:969--972.

\bibitem[Li et~al., 2011]{Li:2011fk}
Li, Y. et~al. (2011).
\newblock Low-coverage sequencing: Implications for design of complex trait
  association studies.
\newblock {\em Genome Res}.

\bibitem[Mardis et~al., 2009]{Mardis:2009qf}
Mardis, E.~R. et~al. (2009).
\newblock Recurring mutations found by sequencing an acute myeloid leukemia
  genome.
\newblock {\em N Engl J Med}, 361:1058--66.

\bibitem[Martin et~al., 2010]{Martin:2010dz}
Martin, E.~R. et~al. (2010).
\newblock {SeqEM}: An adaptive genotype-calling approach for next-generation
  sequencing studies.
\newblock {\em Bioinformatics}.

\bibitem[Nakamura et~al., 2011]{Nakamura:2011kx}
Nakamura, K. et~al. (2011).
\newblock Sequence-specific error profile of illumina sequencers.
\newblock {\em Nucleic Acids Res}.

\bibitem[Nielsen et~al., 2011]{Nielsen:2011fk}
Nielsen, R. et~al. (2011).
\newblock Genotype and {SNP} calling from next-generation sequencing data.
\newblock {\em Nat Rev Genet}, 12:443--51.

\bibitem[Paten et~al., 2008]{Paten:2008uq}
Paten, B. et~al. (2008).
\newblock Enredo and pecan: genome-wide mammalian consistency-based multiple
  alignment with paralogs.
\newblock {\em Genome Res}, 18:1814--28.

\bibitem[Pleasance et~al., 2010a]{Pleasance:2010bh}
Pleasance, E.~D. et~al. (2010a).
\newblock A comprehensive catalogue of somatic mutations from a human cancer
  genome.
\newblock {\em Nature}, 463:191--6.

\bibitem[Pleasance et~al., 2010b]{Pleasance:2010dq}
Pleasance, E.~D. et~al. (2010b).
\newblock A small-cell lung cancer genome with complex signatures of tobacco
  exposure.
\newblock {\em Nature}, 463:184--90.

\bibitem[Roach et~al., 2010]{Roach:2010oq}
Roach, J.~C. et~al. (2010).
\newblock Analysis of genetic inheritance in a family quartet by whole-genome
  sequencing.
\newblock {\em Science}, 328:636--9.

\bibitem[Robison, 2010]{Robison:2010ys}
Robison, K. (2010).
\newblock Application of second-generation sequencing to cancer genomics.
\newblock {\em Brief Bioinform}, 11:524--34.

\bibitem[Schaid et~al., 2002]{Schaid:2002qf}
Schaid, D.~J. et~al. (2002).
\newblock Score tests for association between traits and haplotypes when
  linkage phase is ambiguous.
\newblock {\em Am J Hum Genet}, 70:425--34.

\bibitem[Shah et~al., 2009]{Shah:2009cr}
Shah, S.~P. et~al. (2009).
\newblock Mutational evolution in a lobular breast tumour profiled at single
  nucleotide resolution.
\newblock {\em Nature}, 461:809--13.

\bibitem[Yi et~al., 2010]{Yi:2010zr}
Yi, X. et~al. (2010).
\newblock Sequencing of 50 human exomes reveals adaptation to high altitude.
\newblock {\em Science}, 329:75--8.

\end{thebibliography}

\end{document}